\documentclass[conference]{IEEEtran}

\usepackage{graphicx}
\usepackage{comment}
\usepackage[colorlinks,urlcolor=blue,linkcolor=blue,citecolor=blue]{hyperref}

\usepackage{hyperref}
\hypersetup{ 
 colorlinks=true, 
 linkcolor=red, 
 filecolor=blue, 
 citecolor = blue, 
 urlcolor=blue, 
} 

\begin{document}
\title{Introducing Post-Quantum algorithms in Open RAN interfaces
  \thanks{This work was supported by the grant PID2020-113795RB-C33 funded by 
  MICIU/AEI/10.13039/501100011033 (COMPROMISE project), the grant PID2023-148716OB-C31 
  funded by MCIU/AEI/10.13039/501100011033 (DISCOVERY project). Additionally, it also 
  has been funded by the Galician Regional Government under project ED431B 2024/41 
  (GPC).}}

\author{\IEEEauthorblockN{Pedro Otero-García}
\IEEEauthorblockA{\textit{atlanTTic, Unviersity of Vigo} \\
Vigo, Spain \\
pedro.otero@det.uvigo.es}
\and
\IEEEauthorblockN{Ana Fernández-Vilas}
\IEEEauthorblockA{\textit{atlanTTic, University of Vigo} \\
Vigo, Spain \\
avilas@det.uvigo.es}
\and
\IEEEauthorblockN{Manuel Fernández-Veiga}
\IEEEauthorblockA{\textit{atlanTTic, University of Vigo} \\
Vigo, Spain \\
mveiga@det.uvigo.es}
}

\maketitle

\begin{abstract}
  Nowadays, 5G architecture is characterized by the use of monolithic hardware, where
  the configuration of its elements is completely proprietary for each manufacturer.
  In recent years, as an alternative to this centralized architecture, a new model
  has emerged: the Open Radio Access Network (Open RAN). One of its main features
  has been the split of the Base Band Unit (BBU) into new simpler hardware with
  more specific functions approaching to a more modular model. As a consequence of
  this split, new interfaces appeared to connect these components that need to be
  protected. With the developments in the field of quantum computing, traditional
  protection mechanisms for this kind of interfaces may be deprecated in the near
  future. This security issue motivates this paper, which aims to study how to
  integrate post-quantum cryptography (PQC) mechanisms to current security standards,
  such as IPsec and MACsec. In addition, the proposal is also put into practice
  to compare the performance of traditional mechanisms with PQC implementations.
  This research shows that the new implementation does not reduce the performance
  of the aforementioned standards, while the security is reinforced against
  quantum attacks.
\end{abstract}

\section{Introduction \label{sec:intro}}

The evolution of mobile networks has experienced exponential growth in recent times, 
driven by the increasing demand for connectivity and technological innovation. The 5G 
generation is already changing the way society and machines communicate and interact 
with the digital and industrial environment, as shown by~\cite{Dangi2021} 
or~\cite{Attaran2023}.  In fact, work has already begun on defining the next 6G 
generation. In~\cite{Shuping2020}, the expectations of this new generation are outlined 
by promoting human-machine communication and strengthening the security of previous 
standards. Both 5G and on the horizon 6G, are technologies characterized by ultra-fast 
speeds, low latency and higher connection capacity that are fundamental to enable a wide 
spectrum of applications, from autonomous vehicles~\cite{Garcia2021}), IoT 
networks~\cite{Chettri2019}, and smart cities~\cite{Gohar2021} to Industry 
4.0~\cite{Rao2018} and immersive virtual reality~\cite{Torres2020}).

However, one of the main problems with the traditional 5G architecture is the vendor 
lock-in that hinders interoperability between different manufacturers, increasing the 
cost of developing a 5G network in the conventional way~\cite{Tisinger2022}. 
Furthermore, the European Union concluded in its 2022 analysis~\cite{UE2022} on 5G 
networks that a vendor lock-in approach increases the threat surface with respect to an 
open architecture. Both studies promote as a solution the implementation of Open RAN 
(Open Radio Access Network) as a standard for the open development of traditional 5G 
networks. It should be noted that Open RAN is not only a 5G architecture-dependent 
solution; it is also applicable to 6G~\cite{Polese2023}, and thus avoids a hypothetical 
future vendor lock-in in the sixth generation of mobile technology.

Nevertheless, despite the remarkable progress made in creating Open RAN, there is no 
security standard for the new open interfaces that emerge along with the new framework. 
Currently, different suites of conventional security protocols are used to protect this
kind of interfaces such as MACsec~\cite{802.1AE-2018}, IPsec~\cite{RFC6071} or 
TLS~\cite{RFC5246,RFC8446}, each with its advantages and drawbacks. MACsec, at the data 
link layer, encrypts packets before they leave the device, ideal for local networks. 
IPsec, at the network layer, encrypts each packet individually end-to-end, offering 
greater flexibility for larger networks. TLS, at the transport layer, encrypts 
communication between specific applications, such as web browsers and servers, as part 
of HTTPs (Hyper Text Transport Protocol Secure).

When analyzing security in 5G networks, advances in quantum computing~\cite{Mehic2023} 
should not be overlooked. Quantum computing represents an imminent threat to nowadays 
cybersecurity. Its exponential ability to perform complex computations puts at risk the 
robustness of traditional cryptographic algorithms that protect digital communications.
Faced with this risk, the scientific community is working on the development of new 
cryptographic algorithms resistant to quantum attacks, following two complementary approaches: Post-Quantum Cryptography (PQC) and Quantum Key Distribution (QKD). PQC 
focuses on designing resilient algorithms, but implementable in classical systems. These 
algorithms are usually based on mathematical problems that are not vulnerable to known 
quantum attacks. e.g., Lattice-Based Cryptography~\cite{Micciancio2009}; with some 
examples like Frodo, Kyber or Dilithium, which are based on Learning With Errors (LWE) 
and Shortest Vector Problem (SVP). The two last ones have been selected in the NIST 
standard of post-quantum algorithms~\cite{Alagic2022}. Code-Based 
Cryptography~\cite{Weger2022}; algorithms like Classic McEliece, BIKE or HQC based on 
error-correcting codes have also been selected for round 4 of the above-mentioned NIST 
competition, although they have not been chosen for standardization. Isogeny-Based 
Cryptography~\cite{Naehrig2019}; the best exponent is SIKE whose security comes from the 
isogeny-path problem, which involves finding a path between two related elliptic curves 
by means of an isogeny. SIKE has also been selected for the fourth round of the 
competition.  Finally, Hash-based cryptography~\cite{Srivastava2023} used by signature 
schemes that use hash functions in their procedure. The only algorithm of this family 
selected by NIST to be standardized is SPHINCS+ a stateless signature scheme.

There are already studies on how quantum resilient solutions could be integrated into 5G
networks. Going back to~\cite{Mehic2023}; apart from mentioning the advances in quantum 
computing it is also assured that QKD systems are suitable for protecting static 
segments of 5G networks where it is feasible to connect QKD nodes over a dark optical 
fiber, while PQC mechanisms are envisioned for mobile and dynamic segments. In addition,
it is also mentioned that PQC is expected to be used temporarily to protect certificates 
and establish connections that can then be secured with QKD-generated keys, ensuring 
that the available methods are already mature enough and their development and adoption 
is expected to accelerate with further investments. On the other hand, more practical 
studies are also found in the literature, for example~\cite{Scalise2014} that shows how 
to implement KEM (Key Encapsulation Mechanism) algorithms together with TLS to secure 
gNB connections to the 5G core. In the tests performed, it was observed that the 
integration of KEM maintains a proper QoS (Quality of Service) improving security 
compared to its conventional alternative.


In this paper, we propose a cryptographic solution resistant to quantum computing 
attacks to protect Open RAN open interfaces. This is achieved by integrating post-
quantum cryptography (PQC) into established security protocols such as IPsec, using a 
hybrid approach that combines traditional methods, such as Diffie-Hellman, with post-
quantum key encapsulation mechanisms (KEM). The paper implements and evaluates four KEM 
algorithms (CRYSTALS-Kyber, BIKE, HQC and Frodo) in a simulated Open RAN environment, 
measuring parameters such as througput, jitter, encryption time and memory usage.  In 
addition, a test environment based on open source tools such as srsRAN and Open5GS is 
designed to evaluate the protection of the Open Fronthaul interface with IPsec-PQC, 
emulating key Open RAN components such as the Central Unit (CU), Distributed Unit (DU) 
and Radio Unit (RU). The study identifies that Frodo and BIKE algorithms offer a better
balance between performance and security, while Kyber shows a competitive but variable 
behavior and HQC stands out in encryption speed but with greater temporal stability 
problems.

The rest of the paper is organizaed as follows. Section~\ref{sec:o-ran} describes the 
principles of Open RAN, highlighting its modular architecture, main components and 
advantages such as interoperability, along with the security risks involved.  Current 
security solutions are analyzed in Section~\ref{sec:challenges}, including protocols 
such as MACsec, IPsec and TLS, and their limitations in the face of future threats are 
discussed. The integration of post-quantum key encapsulation (KEM) algorithms into IPsec
to protect Open RAN interfaces is presented in Section~\ref{sec:integration}. The 
testbed implementation is presented in Section~\ref{sec:testing} and the experimental 
performance metrics such as throughput, latency and resource usage are analyzed in 
Section~\ref{sec:results}, identifying advantages and disadvantages of each KEM.  The
article concludes in Section~\ref{sec:conclusions} with the feasibility of the proposed 
solutions and proposes lines of future work, such as exploring the use of quantum key 
distribution (QKD) and the integration of PQC in other protocols.

\section{Principles of Open RAN \label{sec:o-ran}}

\subsection{Background}
In order to comprehend Open RAN, it is helpful to first have a general understanding of
how mobile networks work. The Radio Access Network (RAN) and the Core Network (Core) are 
the two domains that form the mobile, or cellular/wireless, network. The RAN is the last 
connection between a cellular device and the network, it consists in a base station, 
also called Base Band Unit (BBU) and an antenna. The antenna sends and receives signals 
to and from the devices that are after digitized in the BBU who transmits it into the 
Core.  While operators have long been able to use one vendor for their core network and 
another for the RAN, interoperability between RAN equipment from different vendors was 
often overlooked in favor of maximizing overall functionality . Consequently, with 
current solutions, it is challenging to mix vendors for the radio and BBU, which are 
typically sourced from the same supplier. The main purpose of Open RAN is to change the 
current scenario and enable operators to create their networks and choose the 
manufacturers and hardware at their convenience avoiding the current vendor lock-in.

\subsection{Key Open RAN principles}

\begin{figure}[t]
  
    \centerline{\includegraphics[width=0.9\columnwidth]{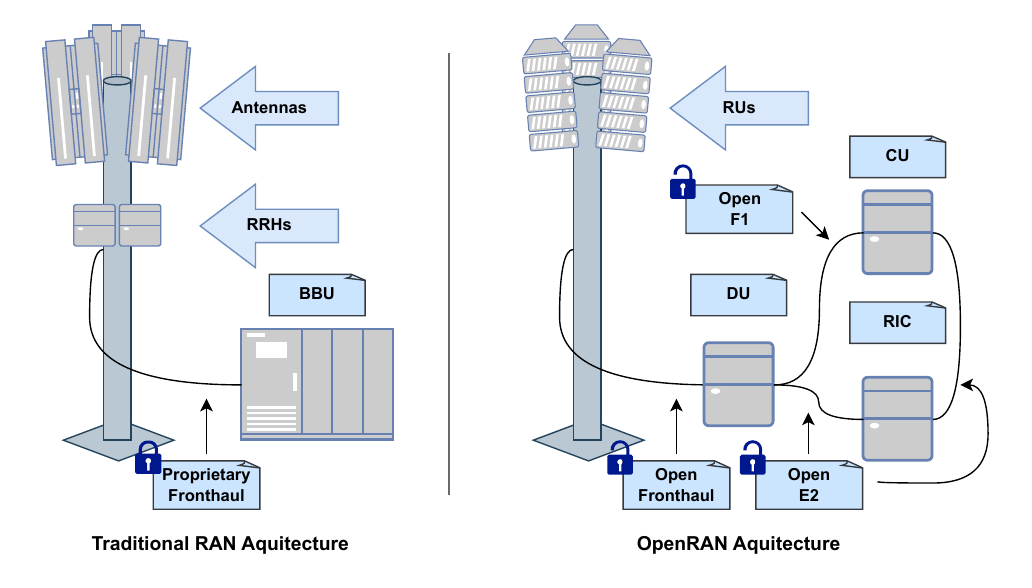}}
    \caption{Traditional RAN topology in front of Open RAN Topology   \label{fig:dissaggregation}}
\end{figure}

Keeping in mind the motivation behind the Open RAN proposal, in this section we
briefly review the main features implemented by this new standard to open and
enhance the traditional RAN.

\begin{itemize}
\item \textbf{Disaggregation}: The Base Band Unit in traditional RAN is the
  equipment in charge of most part of data and signal processing in 5G
  communication, which makes it an expensive and crucial element in the
  architecture. For this reason, the 3GPP defines a split of the BBU into a
  Central Unit (CU) and Distributed Units (DU) to make the RAN management
  more efficient. A visual representation of this change can be seen in
  Figure~\ref{fig:dissaggregation}.

\item \textbf{Intelligence and Automated Control}: Open RAN incorporates
  near-real-time RAN Intelligent Controllers (RICs) and non-real-time RICs to
  optimize network performance across different timescales. These controllers
  utilize xApps and rApps, which, through AI and machine learning algorithms,
  autonomously manage resources and optimize policies.

\item \textbf{Virtualization}: RAN functions and even elements, as for example
  the RICs, can be implemented as software services on cloud platforms,
  reducing dependence on proprietary hardware. This approach facilitates the
  adoption of innovative solutions and enables faster deployment of applications
  and services.

\item \textbf{Open and Standardized Interfaces}: Having a defined method of
  communication among the elements in the RAN, let manufacturers create their
  own components ensuring that they will be able of integrate into any topology
  that follows the open standard.
\end{itemize}

\subsection{Main components of Open RAN}

In the Open RAN framework new components appeared as a result of the
disaggregation, which are interconnected using open interfaces. The most relevant
ones are explained in this section:
\begin{itemize}
\item \textbf{Radio Unit} (RU): Is the corresponding element of the Raidio Head (RU)
  or Remote Radio Unit (RRU) in traditional RAN. RU is responsible for transmitting
  and receiving radio frequency signals and handles the conversion between the
  digital to the analog domain.

\item \textbf{Central Unit} (CU): Is a virtualized network component responsible
  for higher-layer processing functions. Specifically, the CU handles
  non-real-time, packet-based functions such as user plane and control plane protocols.

\item \textbf{Distributed Unit} (DU): It works in layers 1 and 2 of the OSI
  model, among its duties are found error correction and packet scheduling being
  the intermediate point between the RU and the CU.

\item \textbf{RAN Intelligent Controller} (RIC): A logical component that optimizes
  the network in real-time using data and AI algorithms. It is divided into the
  near-RT RIC for millisecond-scale optimization and the non-RT RIC for
  longer-term adjustments within the Service Management and Orchestration
  (SMO) platform.
\end{itemize}

\subsection{Interfaces in Open RAN}

The interfaces in Open RAN facilitate the interaction and control among units shown
in the last Section, for example:
\begin{itemize}
\item \textbf{Open Fronthaul Interface (O-FH)}: This interface connects the DU to
  one or more RUs, handling data flows necessary for RF processing. It is divided
  into four planes: Control, User, Synchronization, and Management.

\item \textbf{F1 interface}:In Open RAN is a crucial link connecting the CU and
  DUs within the network. This interface is key to the disaggregated architecture,
  allowing the CU and DU to operate independently and be sourced from different
  vendors, promoting interoperability and flexibility.

\item O1 Interface: This interface connects the SMO system to network components,
  specifically enabling operations and maintenance functions. 

\item O2 Interface: Connecting the SMO to the O-Cloud (virtualization infrastructure
  supporting Open RAN), the O2 interface allows for resource provisioning and
  workflow management across virtualized network components.

\item A1 Interface: This interface links the non-RT RIC and near-RT RIC to
  exchange policyguidance and strategic control information. Through A1, operators
  can set policies that influence near-real-time optimization functions, making
  the network more adaptive to real-time demands.

\item E2 Interface: The E2 interface connects the near-RT RIC to other RAN elements,
  like the  CU and DUs.
\end{itemize}

\section{Security Challenges and Current Solutions for Open RAN Interfaces\label{sec:challenges}}

The open and modular nature of OpenRAN, while providing significant advantages in
terms of interoperability and flexibility over vendors, introduces potential
security challenges that are either not present in traditional RAN architectures,
or are significantly less risky. As discussed in~\cite{Liyanage2023}, these
challenges arise primarily from the disaggregation of network components and the
use of open interfaces, leading to a larger attack surface and increasing the
complexity of the system.

The disaggregation of the Base Band Unit (BBU), into a Distributed Unit (DUs) and
a Central Unit (CUs), and the addition of elements such as de RIC results in a
greater number of interconnected components that can be used as potential entry
point for attackers. A breach in one component could jeopardize the security of
the entire network. For instance, attackers could perform Man-in-the-Middle (MitM)
attacks by intercepting communications between the DU and CU; allowing them to eavesdrop 
on sensitive data, alter communications, or redirect traffic.
Furthermore, vulnerabilities in individual units may lead to device hijacking,
where attackers take control of the unit disrupting network operations or
facilitating new attacks on connected systems.

Open RAN encourages the incorporation of new vendors and the combination of the
different units they offer with those of other suppliers. However, the integration
of components from different vendors, as long as there is no proven communication 
standard in the open interfaces, can lead to security breaches when transmitting
data between the different units. Attackers can exploit these inconsistencies to
exploit interoperability attacks, in which weaknesses in inter-vendor communication
protocols are used to inject malicious traffic or to interrupt services. In
addition, reliance on third-party hardware and software increases the risk of
supply chain attacks, where compromised components or backdoors are introduced
during manufacturing or integration, leading to long-term security vulnerabilities.

The adoption of open interfaces also presents significant risks compared to
proprietary interfaces. Unlike proprietary protocols, which often require
specialized knowledge or can only be configured by employess from the manufacturer,
communication over open interfaces is public and documented. This scenario makes
malicious actors more interested in exploiting this type of systems because it is
easier to identify weak points. Common attacks on open networks include the
forwarding of previously captured data packets to bypass authentication or disrupt
system functionality, or packet injection attacks, in which malicious packets are
manipulated to alter control signals or introduce false information into the
network. Open interfaces also make Open RAN systems particularly susceptible
to denial-of-service (DoS) attacks, in which attackers flood the interfaces
with traffic to degrade or completely disrupt network operation.

It is also worth remembering that within Open RAN the integration of Artificial
Intelligence (AI) and Machine Learning (ML) is being developed and implemented in
Open RAN components, most notably in the RAN Intelligent Controller (RIC). These
systems, designed to optimize network performance, can be targeted by adversarial
AI attacks. For example, attackers could poison network data used to train AI/ML
models, resulting in incorrect decisions that would degrade quality of service
(QoS), misallocate resources, or even cause hardware or software failures.

To address these challenges, Open RAN frameworks currently rely on a range of
established security suites of protocol, including MACsec, IPsec, and TLS. Each
of these suites offers specific advantages and limitations, making them suitable
for different types of Open RAN interfaces depending on the performance and
security requirements.

MACsec operates at the data link layer, is highly effective for protecting local,
high-speed connections such as the Open Fronthaul (O-FH) interface, which connects
the DU and the RU. Its lightweight encryption ensures minimal impact on latency,
making it ideal for real-time communication. Nonetheless, its scope is limited
to localized networks and does not provide end-to-end encryption, making it
unsuitable for broader protection needs.

IPsec, on the other hand, provides robust end-to-end encryption at the network
layer. Its flexibility and configurability make it well-suited for securing
interfaces like E2, which handles critical control data between the near-real-time
RIC and other components. Despite its strengths, IPsec introduces additional
overhead due to the extra headers required for its encryption protocols, which
can impact efficiency in bandwidth-constrained scenarios.

Finally, TLS, a widely adopted transport-layer protocol, is used primarily to
secure application-level communications and management data. Its streamlined
handshake process, particularly in TLS 1.3, ensures fast and efficient secure connection 
establishment. While it is effective for interfaces such as O1, O2,
and A1, which handle less latency-sensitive management and policy data, TLS is
less suitable for real-time applications and lacks the multi-layer security coverage
of protocols like IPsec.

Although these protocols provide a strong foundation for securing Open RAN, no
single solution is universally optimal. The selection of a security protocol depends
on the specific interface and its requirements in terms of latency, throughput,
and scope of encryption. Furthermore, as quantum computing progresses, it is
critical to explore cryptographic solutions that are resistant to quantum-based
attacks, which threaten to undermine many of the encryption methods used today.
This ongoing need for evolution in security mechanisms highlights the importance
of developing future-proof protocols for Open RAN systems.

\section{Integrating PQC into the open interfaces of Open RAN \label{sec:integration}}

Traditional cryptography is responsible for ensuring secure communication and
data protection across different information channels; providing confidentiality,
integrity, and non-repudiation. However, with the constant progress in quantum
computing (QC), security in classical cryptography can be compromised in the
near future. QC was first conceptualized  in~\cite{feynman82}; it leverages
principles of quantum mechanics,  such as superposition and entanglement, to
achieve computational capabilities beyond classical systems.

Classic cryptography is usually split in two kind of encryption: symmetric and
asymmetric encryption. Symmetric encryption algorithms, such as AES (Advanced
Encryption Standard), use  a shared secret key for encryption and decryption.
The security of these systems depends on the size of the key space, making
brute-force attacks computationally infeasible for  sufficiently large keys. 
Nevertheless, Grover's algorithm~\cite{Grover1996}, a quantum search algorithm,
reduces the effective key space exponentially by half, significantly weakening
their security. I.e., when an encryption scheme use AES-128 as its encryption
algorithm in a quantum system, the real security achieved will be equivalent
as the one in AES-64 in a classic system~\cite{mavroeidis2018}.

Regarding asymmetric cryptography, it relies on one-way mathematical functions,
such as integer factorization (RSA) or the discrete logarithm problem
(Diffie-Hellman and Elliptic Curve Cryptography, ECC). These algorithms are the
base of secure communications and digital signatures in the current information
systems. The main thread of asymmetric encryption is Shor's quantum algorithm that
can efficiently solve these problems, rendering such cryptosystems insecure.

Currently, the research community is developing quantum and post-quantum
cryptographic solutions to face the hazards of algorithms such as Shor's and
Grover's. On the one hand, Quantum Key Distribution (QKD) protocols are a secure
communication method that leverages quantum mechanics to distribute quantum-safe
encryption keys between parties. Protocols such as BB84 or COW~\cite{nurhadi2018}
are examples of QKD implementations. On the other hand, post-quantum cryptography
(PQC) can be developed in traditional computers and still be resistant to quantum
attacks. An example of PQC algorithms are the Key Encapsulation Mechanisms
(KEM)~\cite{alagic2019}, they are cryptographic primitives that enables a sender
to securely generate and transmit a short, random secret key to a receiver,
without revealing the key to an eavesdropping adversary.

Since the Open RAN framework is emerging as the standard model for the future of 
communication networks, it is of vital importance to formalize the security of its
open interfaces, as detailed in Section 2. This approach seeks to ensure that the 
interoperability promoted by Open RAN does not compromise the robustness of 
communications in the face of modern and emerging threats. However, the security 
protocol suites currently employed base their reliability on classical algorithms, which 
are vulnerable to quantum computing attacks already identified (see previous section).

In this context, the present work aims to propose a quantum computing resistant solution 
to protect some of the Open RAN open interfaces. The cryptography
selected for implementation is post-quantum, due to its ease of integration
in conventional computers, allowing a more accessible transition to network
operators. At the same time, quantum cryptography based on key distribution (QKD)
is left for future research, since its deployment requires additional infrastructure
and currently presents significant practical barriers.

\subsection{Implementing PQC in IPsec}

IPsec has been chosen as the base protocol suite for protecting open interfaces
due to its maturity, flexibility and wide adoption in secure networks. It should
be noted that in the absence of a universal standard for Open RAN interface
protection, no protocol suite is optimal for all scenarios, highlighting the
need for research such as the present one.

\begin{figure}[t]

    \centerline{\includegraphics[width=24pc]{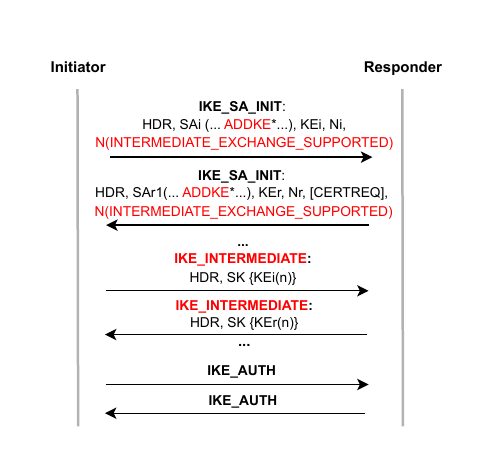}}
    \caption{IKEv2 with intermediate messages following RFC 9370.    \label{fig:rfc9370}}
\end{figure}

One of the strategies proposed to ensure the security of communications when
integrating post-quantum cryptography (PQC) into IPsec is to adopt a hybrid
approach. This approach combines classic key exchange like Diffie-Hellman (DH)
or Eliptic Curve Diffie-Hellman (ECDH), used in the IKEv2 phase of IPsec, with
one or more PQC algorithms of Key Encapsulation Mechanisms (KEMs). The
implementation of multiple key exchange in IKEv2 is described in RFC 
9370~\cite{RFC9370}. Basically, when the initiator sends the \textsc{ike\_sa\_init}
adds a notification payload \textsc{intermediate\_exchange\_supported} indicating
the possibility of using multiple key exchanges and one Additional Key Exchange
Transform Type (\textsc{addke}) for each proposed KEM. Then, when the receptor
responds to the \textsc{ike\_sa\_init} with the selected proposals, the
initiator starts the message exchange for each KEM using the \textsc{ike\_intermediate}
messages as can be shown in Figure~\ref{fig:rfc9370}. All the shared keys from
the different algorithms selected will be a combine as a input to a pseudo-random
function that will compute the shared secret used as seed for the encryption
process.

At the time of writing this article, post-quantum key exchange algorithm
implementations are not yet widespread in real cryptographic applications comparing
it with classic cryptographic solutions such as DH or Elliptic Curve DH. It is
also claimed that pre-quantum algorithms are extensively studied and have proof
of their security properties~\cite{giron2023}. Post-quantum security, on the other
hand, is still under scrutiny. Thus, it is not unreasonable to think of scenarios
where a flaw in a post-quantum algorithm is discovered. A hybrid post-
quantum/conventional approach can overcome this challenge, as it ensures
the system is at least as robust as the security provided by the individual
primitives remaining safe against classical attackers.

Furthermore, using a hybrid approach allows current networks to adopt
post-quantum technologies gradually, providing a safe transition period
while definitive standards and best practices are developed.

\subsection{Selection of KEMs for testing}

The National Institute of Standard and Technology (NIST) organized a
post-quantum algorithm standardization process, similar to the one carried out
between 1997 and 2000 to standardize the AES symmetric encryption algorithm or
between 2007 and 2012 for the SHA-3 hash algorithm. In this case, NIST effort
sought to identify algorithms that are not only resistant to quantum attacks
but also have adequate performance in real-world environments, considering both
latency and resource consumption.

The round 4 of the process concluded in 2022~\cite{Alagic2022}, selecting
CRYSTAL-kyber as public-key encryption and CRYSTALS–Dilithium, FALCON, and SPHINCS+
as digital signatures standards. There are already projects such as Open Quantum
Safe~\cite{OQS} (OQS) that implement the participating algorithms, not only the
selected for standardization, allowing their evaluation in practical scenarios.
These developments are essential to identify compatibility, scalability and
performance issues in real environments before mass adoption.

For the integration of PQC into IPsec in the context of this paper, it has been
used the implementations of KEM algorithms available in the open source library
\texttt{liboqs}, developed by OQS. The algorithms selected for testing have been: 
CRYSTALS-Kyber, BIKE, HQC and FRODO. The first three participated in the fourth
round of the NIST standardization process, while FRODO was eliminated after the
third round~\cite{round4}. However, the early elimination of FRODO by NIST in
the process does not imply that it is an unsafe algorithm; indeed is one of the
two post-quantum algorithms recommended by the German Federal Office for
Information Security (BSI) as cryptographically suitable for long-term
confidentiality along with Classic McEliece.

\section{Testing Scenario \label{sec:testing}}

The incorporation of post-quantum cryptographic algorithms into IPsec
introduces important considerations for compatibility, migration, and future
enhancements. Compatibility is largely achieved through hybrid key exchange
mechanisms, combining classical methods such as ECDH with post-quantum KEMs.
This approach guarantees that updated systems can interoperate with legacy ones
during a gradual transition to fully post-quantum deployments. Nonetheless, the
message overhead of post-quantum algorithms during key exchanges, may impact latency-
sensitive applications or low-power devices, requiring careful
optimization. 

Concurrently, integrating post-quantum cryptography with modern protocols like
QUIC presents a promising opportunity for enhancing security in latency-sensitive
applications. The low-latency design of QUIC, combined with the robust encryption
provided by PQC algorithms, could potentially offer a future-proof solution for securing 
modern communications. 

In order to evaluate the performance of the proposed solution, a simple testbed
of an Open RAN environment was designed using different open-source tools
implemented on Linux operating system machines with \texttt{Ubuntu 22.04}
operating system.

\subsection{Test Scenario Topology}

\begin{figure}[t]
  
    \centerline{\includegraphics[width=16pc]{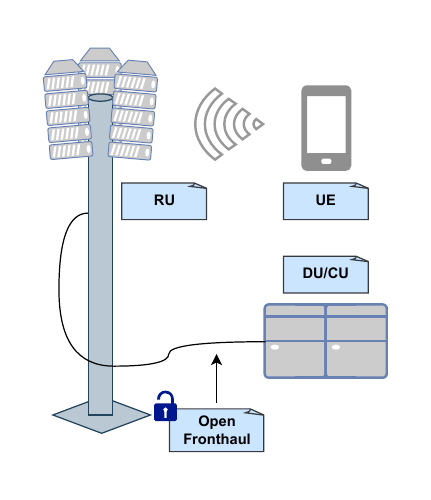}}
    \caption{Topology of the testing scenario.  \label{fig:topology}}
\end{figure}

Figure~\ref{fig:topology}  illustrates the components that comprise the testbed,
which are emulated on different machines:
\begin{itemize}
\item \textbf{User Equipment} (UE): It represents the connections that a terminal
  would have with a 5G network.

\item \textbf{Radio Unit} (RU): It will work as the RAN connection point for the UE.

\item \textbf{Central Unit }(CU) and \textbf{Distributed Unit} (DU): In this
  scenario, both are implemented on the same machine to simplify the topology.
\end{itemize}

To provide Open RAN and Core 5G functionality, the open source projects 
srsRAN~\cite{srsran} and Open5GS~\cite{open5gs} were used as main developing
sources. As detailed in the Figure, the interface selected for testing is the
Open Fronthaul (O-FH). Although in real environments it is preferred to protect
the O-FH using MACsec due to its fast implementation, scientific 
literature~\cite{abdalla2024}\cite{cho2019} has shown that it is feasible to
use IPsec to protect the O-FH interface without compromising the QoS. In this work,
the latter solution was chosen to evaluate its feasibility in terms of
performance\footnote{Although the tests were performed specifically on the O-FH
interface, the protection approach using IPsec combined with PQC is applicable
to other interfaces, such as the E2 interface.}.

\subsection{IPsec Implementation}

Regarding IPsec implementation between RU and DU, the open source library
\texttt{strongSwan}~\cite{strongswan} has been chosen as a widely used
implementation to develop IPsec tunnels that guarantees compliance with IETF
standards for IPsec (RFC 6071~\cite{RFC6071}) and IKEv2 (RFC 7296~\cite{RFC7296}),
ensuring interoperability with different systems and devices. In addition, it
offers support for advanced encryption algorithms, including KEM in beta
version \texttt{6.0.0} integrated with the \texttt{liboqs} library following
RFC 9370~\cite{RFC9370}.

\subsection{Tests Performed}

The objective of the tests is to assess the performance of communication in the
event that PQC-IPsec protects the O-FH interface. Specifically, the following
parameters were evaluated:
\begin{itemize}
\item \textbf{Throughput}: Evaluates the ability of the network to transmit
  data efficiently. This indicator is crucial to analyze the maximum
  achievable throughput in the O-FH interface.

\item \textbf{Delay}: Measures the latency time between sending and receiving
  data, thereby providing valuable information regarding the capacity of the
  network to operate in real time.

\item \textbf{Jitter}: Determines the variation of the delay,, a critical
  factor for delay-sensitive applications.

\item \textbf{Memory Usage}: Analyzes the impact of IPsec encryption with
  PQC implementation on system memory resources.

\item \textbf{Encryption Time}: Measures the time required to encrypt packets
  during communication. This parameter provides insight into the computational
  overhead introduced by the hybrid approach.
\end{itemize}

\section{Performance Results\label{sec:results}}

This section provides a detailed analysis of the results obtained from measuring
the performance variables outlined in the previous section

\begin{figure}[t]
    \centering
    \includegraphics[width=0.45\textwidth]{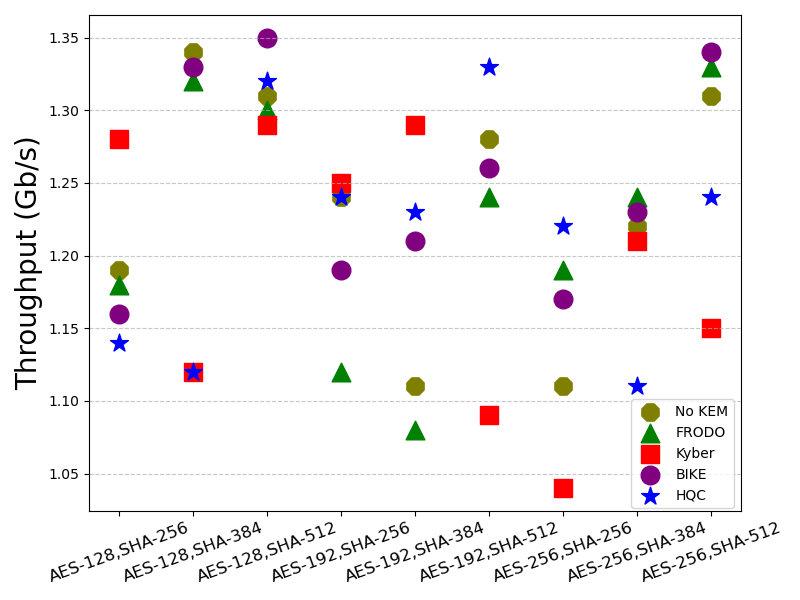}
    \includegraphics[width=0.45\textwidth]{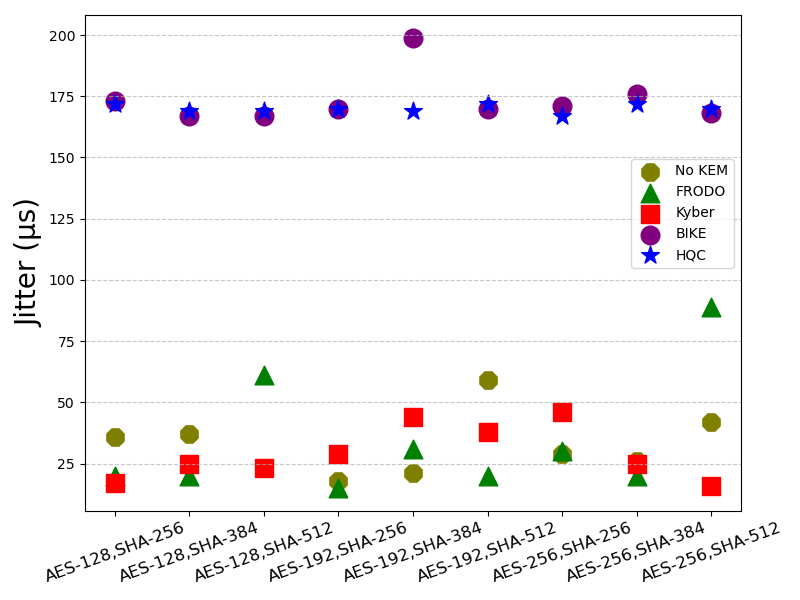}
    \caption{\label{fig:througput} Throughput and jitter comparison between KEM  algorithms.}
\end{figure}

Figure~\ref{fig:througput} reflects the throughput, whereby higher values
indicate better performance. In this context, the incorporation of KEM into the
IKEv2 protocol has a different impact depending on the scheme used. Frodo maintains
a throughput very close to the case without KEM, which makes it an efficient and
stable option. BIKE also shows consistent behavior, while HQC achieves good results
in specific configurations but introduces more variability. Kyber, while
showing competitive values in some cases, has more erratic behavior. Configurations
with AES-128 tend to maintain a high and stable throughput, while AES-192 and
AES-256 show greater fluctuations. KEM selection should consider not only security,
but also its impact on throughput, with Frodo and BIKE being the options with best
performance in this test. The jitter results, also in Figure~\ref{fig:througput},
reflect that the addition of KEM to the IKEv2 protocol significantly affects the
temporal stability of transmissions. Without KEM, the values are moderately low
but variable depending on the configuration. Frodo improves jitter slightly in
several configurations, such as AES-128/192,SHA-256 (20 $\mu$s/ 15 $\mu$s, although
in others, such as AES-256,SHA-512 (89 $\mu$s), it introduces more jitter.
Kyber also achieves good performance in general, standing out in configurations
such as AES-128/256,SHA-256 (16 $\mu$s, 17 $\mu$s), with less consistent values
in configurations such as AES-192,SHA-384 (44 $\mu$s). On the other hand, BIKE and
HQC generate consistently high jitter (range between 167 and 199 $\mu$s),
making them less suitable for time-stability sensitive applications.
Frodo and Kyber are the most favorable choices, especially in configurations
optimized for real-time application like VoIP or video calls.

Concerning the results regarding encryption time of the packets, it is shown that
without KEM the times vary between 5 and 11 $\mu$s. Frodo significantly improves
performance, with lower times (3 to 6 $\mu$s) in most configurations. Kyber has
mixed performance, improving in some configurations (e.g., 4 $\mu$s on
AES-256,SHA-512), but getting worse in others (10 $\mu$s on AES-192,SHA-384).
BIKE and HQC are the fastest, with very low times (2 to 4 $\mu$s), standing out
especially in configurations such as AES-256,SHA-256. In summary, BIKE and HQC
offer the best speed, followed by Frodo, while Kyber has more variable performance.

\begin{figure}[t]
    \centering
    \includegraphics[width=0.445\textwidth]{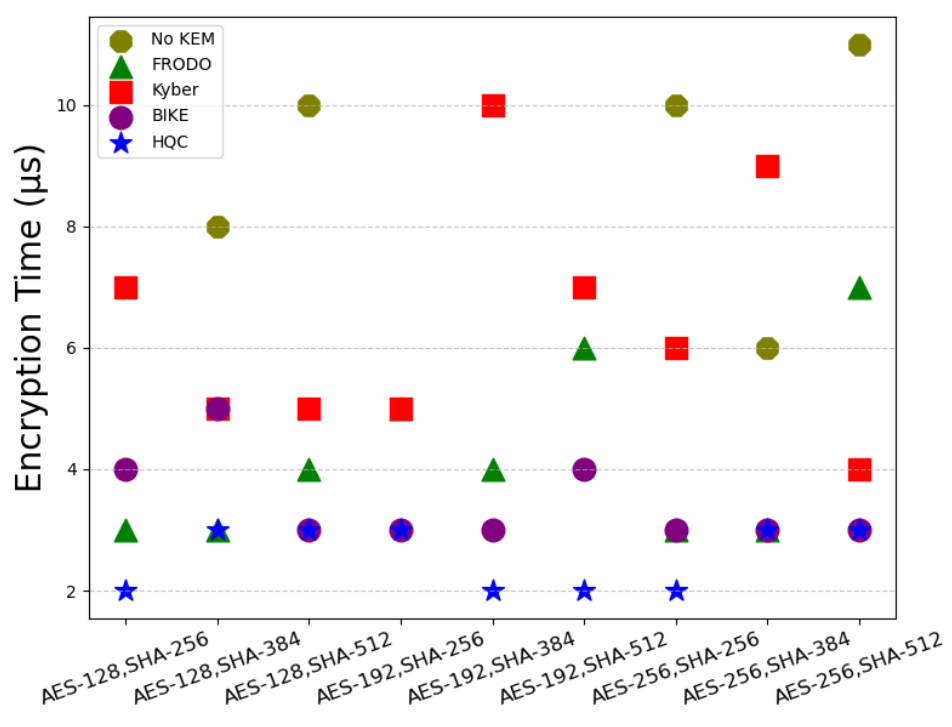}
    \includegraphics[width=0.45\textwidth]{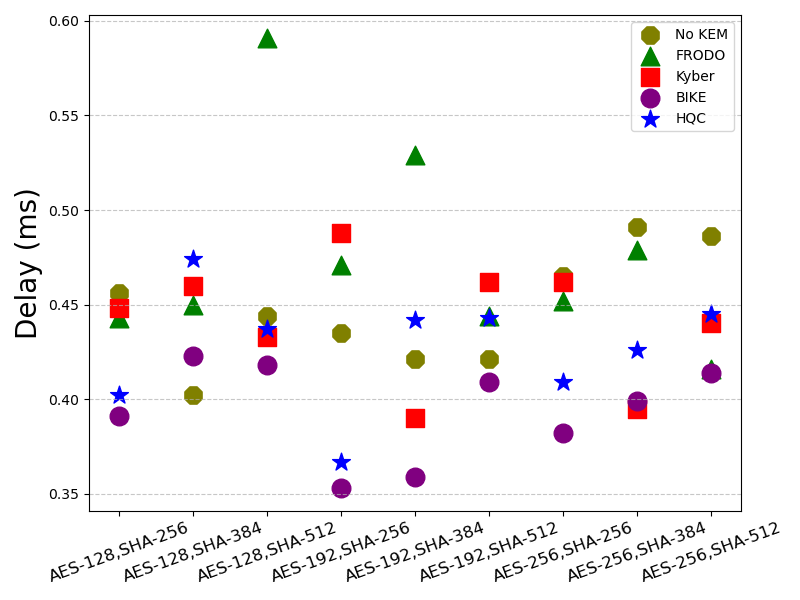}
    \caption{\label{fig:enc_time}Encryption time comparison between KEM algorithms.}
\end{figure}

As for the delay, it shows how the incorporation of KEM affects the latency in
sending and receiving packets. Without KEM, the delay values are relatively
low, varying between $0.402$ and $0.491$ ms. Frodo, while improving in some cases
such as AES-128,SHA-256 ($0.443$ ms) and AES-256,SHA-512 (0.416 ms), introduces
a noticeable increase in configurations such as AES-128,SHA-512 (0.591 ms).
Kyber presents a more stable performance, with values between $0.39$ and $0.488$
seconds, showing a moderate but consistent delay in all configurations. BIKE has
the best results, with the lowest delay in several configurations, standing out
in AES-192,SHA-256 ($0.353$) and AES-192,SHA-384 ($0.359$). HQC, on the other
hand, maintains relatively low delay values, but not as optimized as BIKE, with
a range of $0.367$ to $0.474$ ms. That is, BIKE presents the lowest delay,
followed by Kyber and Frodo, while HQC remains competitive, but with slightly
more latency in comparison.

\begin{figure}[t]
 
    \centerline{\includegraphics[width=20pc]{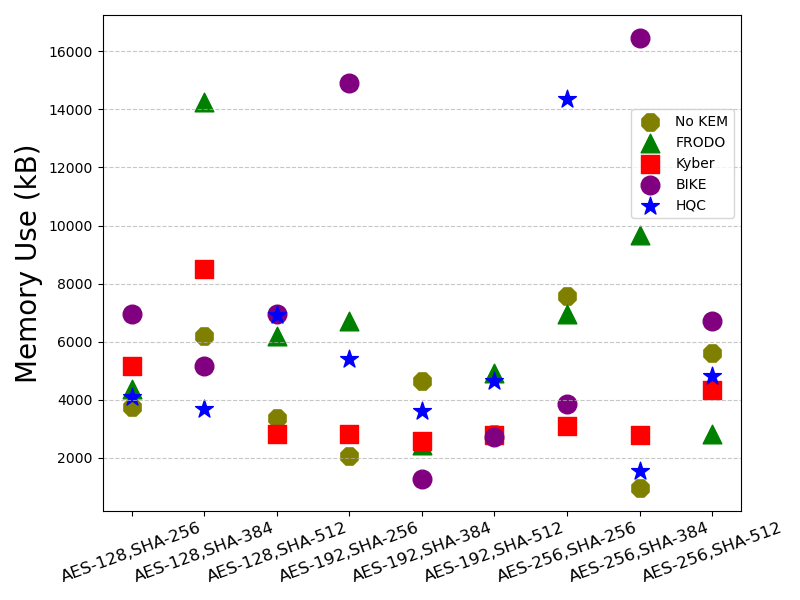}}
    \caption{Memory usage by the different KEM proposals.   \label{fig:mem_use}}
\end{figure}

The memory usage results, Figure~\ref{fig:mem_use} reveals distinct differences
among the KEM schemes. Without KEM, memory usage is relatively low, ranging from
950 kB to 3,750 kB. Frodo introduces significant memory overhead, especially
in configurations like AES-128,SHA-384 (14,250 kB) and AES-256,SHA-512 (9,675 kB).
Kyber shows more moderate memory increases, ranging from 2,820 kB to 5,160 kB.
BIKE also results in higher memory usage, especially in AES-128,SHA-512 (14,900 kB)
and AES-256,SHA-384 (16,475 kB). HQC presents a more consistent memory demand,
ranging from 3,600.384 kB to 6,925 kB. Overall; Kyber, followed by HQC, has the
best performance in terms of memory usage efficiency compared to the KEM-less version.

\section{Conclusions \label{sec:conclusions}}

This article has highlighted the importance of protecting the open
interfaces  of Open RAN even in the case of the imminent threats of
quantum computing. As possible solutions within the study conducted, we have
proposed the implementation of IPsec together with post-quantum key
encapsulation algorithms such as Kyber, Frodo, HQC or BIKE. 

In order to demonstrate the feasibility of the proposed solution, a
testbed scenario has been designed based on open-source projects, mainly srsRAN
and Open5GS. Within this setting, we measured Throughput, Delay, Jitter,
Encryption Time and Memory Usage for different KEMs in combination with
different encryption and integrity algorithms in IPsec. No single combination
has been found to be clearly advantageous in all areas compared to the others.
Frodo and BIKE stand out as the most balanced options. Frodo maintains a high
and stable throughput, comparable to the case without KEM, and offers moderate
improvements  in jitter and encryption time, making it ideal for applications
requiring stability.  BIKE excels in encryption speed and delay, consistently
achieving the fastest times and lowest latency compared to other KEMs, making
it optimal for time-sensitive  applications. Kyber obtains competitive performance
in some configurations, but has shown great variability in the results obtained
being more unstable in the tests. HQC, while achieving excellent encryption speeds
and competitive delays, has higher jitter and variability in throughput, making it
less suitable for applications that require temporal stability. Configurations
with AES-128 offer greater stability in general, while AES-192 and AES-256 tend
to generate more jitter. Again, in the tests performed there is no clearly
better combination, the results are indicative and could differ in a real
non-simulated scenario.

A natural, clear extension to this work is the operational integration
of IPsec-QKD in the Open RAN framework using keys generated with real quantum
devices. Our study will probably focus on the E2 interface, since it is considered
a more realistic use case  than O-FH, after the results obtained in this article.
Besides, there are other protection mechanisms that can be used to protect Open RAN open 
interfaces: MACsec and TLS. We are currently working on the implementation of
PQC and QKD to the mentioned protocol suites to see their viability in the
different Open RAN interfaces.

\end{document}